\documentclass[doublecol]{epl2} 
\usepackage{graphicx}
\usepackage{psfrag}
\usepackage{color}
\usepackage{float}
\usepackage{amsmath}
\usepackage{amsthm}  
\usepackage{amsfonts}
\usepackage{amssymb}
\usepackage{mathrsfs} 
\usepackage{fancyhdr}
\usepackage{hyperref}
\usepackage{bm}

\newcommand{\beq}{\begin{equation}}
\newcommand{\eeq}{\end{equation}}
\newcommand{\ben}{\begin{eqnarray}}
\newcommand{\een}{\end{eqnarray}}
\newcommand{\baq}{\begin{array}}
\newcommand{\eaq}{\end{array}}

\newcommand{\eef}{\end{figure}}

\title{ {Subradiant hybrid states} in the open 3D Anderson-Dicke model}
\shorttitle{Subradiant hybrid states in the open 3D Anderson-Dicke model} %Insert here a short version of the title if it exceeds 70 characters

\author{A. Biella\inst{1} \and F. Borgonovi\inst{1,2,3} \and R. Kaiser\inst{4} \and G. L.  Celardo\inst{1,2,3}  }
\shortauthor{A. Biella \etal}

\institute{                    
  \inst{1} Dipartimento di Matematica e Fisica,
 Universit\`a Cattolica, via Musei 41, 25121 Brescia, Italy\\
  \inst{2} Interdisciplinary Laboratories for Advanced Materials Physics,
 via Musei 41, 25121 Brescia, Italy\\
 \inst{3} Istituto Nazionale di Fisica Nucleare,  Sezione di Pavia, 
via Bassi 6, I-27100,  Pavia, Italy\\
 \inst{4} Universit\'e de Nice Sophia Antipolis, CNRS, Institut Non-Lin\'eaire de Nice, UMR 7335, Valbonne F-06560, France
}

\abstract{Anderson localization is a paradigmatic coherence effect in disordered systems, often analyzed in the absence of dissipation.
Here we consider the case of coherent dissipation, occurring for open system with coupling to a common decay channel.  
This dissipation induces cooperative Dicke super- and subradiance and  
an effective long range hopping, expected to destroy Anderson localization.  
We are thus in presence of two
competing effects, i.e localization driven by disorder 
and delocalization driven by
dissipative opening. 
Here we demonstrate the existence of  a  {{\it subradiant hybrid regime}}, 
emerging from the interplay of opening and
disorder, in which subradiant states are  {hybrid with both features of localized 
and extended states}, while superradiant states are extended. 
We also provide analytical predictions for this regime, confirmed by numerical simulations. }

\begin{document}

\maketitle

\section{Introduction}

One of the most interesting effect induced by quantum coherence
is Anderson localization~\cite{Anderson}. 
This effect  is relevant for many areas of Physics,
quantum computing~\cite{qc}, 
cold atoms~\cite{kaiser}, 
light harvesting systems~\cite{fmo}, 
mesoscopic devices~\cite{Beenakker,newprl}.
Initial search of experimental evidence of Anderson localization 
for non interacting waves
in 3 dimensions has been limited by the presence of absorption ~\cite{Lagendijk97}.
Recent progress has however been obtained in a large variety of fields 
of research, including 
acustic waves~\cite{skynat}, light waves~\cite{maret},
matter waves~\cite{Garreau}.

Whereas absorption and dissipation are often considered to be limiting phenomena for 
observing Anderson localization, the situation of opening with corresponding 
 coherent dissipation has not been systematically addressed. 
This situation has been
 considered in a different research community \cite{fmo,Friedberg}, investing cooperative effect induced by coupling to a commun decay
 channel, following the pioneering work by Dicke in 1954~\cite{dicke54}.
The question whether localization can survive 
coherent dissipation in open $3D$ systems has been discussed
 in particular in the case of 
resonant light scattering ~\cite{Orlowski, kaiser}.  

In this letter we focus on the case of coherent dissipation due
to the coupling to a common decay channel,
relevant for many realistic situations~\cite{kaiser,lagendijk,newprl}.
If this coupling is strong enough,  dissipation induces a well known 
coherent effect: Dicke super- and subradiance~\cite{dicke54},
which induces the so-called Superradiant Transition (ST),
 driven by the opening, 
i.e.  by the coupling to an external
environment characterized by a continuum of states. 
%Note that while Dicke~\cite{dicke54} also considered the regime of 
%many excitations, 
%here we will only deal with the single excitation limit. 
This transition should  be compared on the other hand with the 
Anderson localization, driven by intrinsic local disorder which 
consists in the suppression
of diffusion due to exponential localization  of 
 the eigenfunctions of the system.

Both Dicke superradiance and Anderson localization have been 
widely studied in literature in a separate way, and
their interplay has been poorly analyzed.
In order to highlight the fundamental aspects of this interplay, 
we  
  {study} a simple but
general model: the $3D$  Anderson model~\cite{Anderson}.
A related $1D$ model, without a
metal-insulator transition, has been already considered  in
Ref.~\cite{alberto}.  
A common feature of disordered systems with coherent dissipation  is
the competition between long range hopping induced by opening
and localization induced by disorder. 
It is commonly accepted that any kind of long range
hopping, decaying slower than $1/r^d$, where $d$ is the system
dimension, destroys localization~\cite{Anderson}.  {On the contrary, here  we show} that
the very correlated nature of long range hopping due to the opening
induces a Superradiant Transition which allows to
preserve  {some feature of localization}. 

Dicke superradiance occurs in the large opening regime and 
it cannot be treated by perturbation theory.
In such a case the effective non-Hermitian Hamiltonian approach to open quantum
systems has been shown to be  very effective~\cite{heff}.
Non-Hermitian Hamiltonians have been already employed 
in random matrix theory~\cite{verbaarschot85,puebla,jung1,sokolov},
in paradigmatic models
of coherent quantum transport~\cite{kaplan,rottertb} 
and in realistic open 
quantum systems~\cite{qdots,kaiser,zelenuclear,lagendijk,rotter2}.

\section{The Model}
In the closed $3D$  Anderson model, a particle
hops between  neighbors sites of a $3D$ cubic lattice 
with $N$ sites, in the presence of 
on--site disorder. 
The Hamiltonian of the closed $3D$ Anderson model can be written as: 
\begin{equation}
H_0= \sum_{j=1}^{N} E_j | j\rangle \langle j| + \Omega \sum_{\langle i,j\rangle} \left(| j \rangle \langle i|
+| i \rangle \langle j|\right) \,,
\label{AM}
\end{equation}
where the summation $\langle i,j\rangle$ runs over 
the nearest-neighbor sites, $E_j$ are random variables uniformly distributed
in $[-W/2 ,+W/2]$,  $W$ is a disorder parameter,
 and $\Omega$ is the tunneling transition amplitude.
%(in our numerical simulations we set $\Omega=1$).
The nature of the eigenstates of the $3D$ Anderson model
depends on the degree of disorder: 
for small disorder 
the states in the middle of the energy band are extended,
while close to the band edges, for energies below the
\emph{mobility edges}, they are localized~\cite{sheng}.
On increasing $W/\Omega$, the mobility edges approach one to each other 
and above the  critical value $W / \Omega  {\simeq} 16.5$~\cite{sheng}, 
all states become localized and a global AT occurs.
In the localized regime the shape of  eigenfunctions behaves as
$|\psi(j)| \sim \exp(-|\vec{x_j}-\vec{x_0}|/\xi)$,  
where $\vec{x_j}$ are position vectors and  $\xi$ is the
localization length.
In this work we will disregard the effect of the mobility edges, 
focusing on the global AT:
\begin{equation}
AT \quad  {\rm at} \quad  W /  \Omega {\simeq}16.5 
\label{AT}
\end{equation}

Our main question is about the effect of opening on the  {localization properties of the eigenstates}. 
We open the $3D$ Anderson model by allowing
the particle to escape the system from any site into the same continuum
channel. This situation of ``coherent dissipation'', can be met
in many realistic systems~\cite{kaiser,newprl},
% mesoscopic
%devises~\cite{exp} and solid state devices for quantum processing~\cite{newprl}
when the wavelength of the particle 
in the continuum channel is comparable with the sample size.
The case of only one channel in the continuum
is  somehow extreme, 
but we expect that our findings have general validity
whenever more states compete to decay in the same channel.
This kind of opening is different from a standard coupling to a thermal
bath (for instance described by time dependent diagonal terms
in the Hamiltonian~\cite{petruccioni}), and takes into account only the 
particle  escape from the system (dissipation).
The open system is described by the 
effective non-Hermitian Hamiltonian~\cite{alberto}: 
\begin{equation}
(H_{\mathrm{eff}})_{ij}=(H_0)_{ij} -\frac{i}{2}  \sum_c A_i^c (A_j^c)^* = (H_0)_{ij} -i\frac\gamma2 Q_{i,j},
\label{amef}
\end{equation}
where $H_0$ is the Anderson Hamiltonian, Eq.~(\ref{AM}),
$A_i^c$ are the transition amplitudes
from the discrete states $i$ to the continuum channels $c$.
In our case we have one decay channel, $c=1$,  equal
 couplings (
$A_i^1= \sqrt{\gamma}$) so that
  $Q_{ij}=1$ $\forall i,j=1,..,N$ is a full matrix.
The complex eigenvalues of $H_{\rm eff}$ can be written as,
$E_r -i \frac{\Gamma_r}{2}$,
where $\Gamma_r$ are the decay widths of the states. 
Since the average width is given by 
$\gamma$~\cite{heff}, 
the degree of resonance overlap is determined by  
$\kappa=\gamma/D$,
where $D$ is the mean level spacing of the energy levels of $H_0$.
We can regard $\kappa$ as   an  effective degree of opening
and at $ \kappa \simeq 1$~\cite{puebla,heff}
a segregation occurs, i.e.
almost the entire  decay width, $N\gamma$, is allocated to just one 
short-lived ``superradiant'' state, 
while all other $N-1$ states become "subradiant'' with a small decay width.
We refer to this segregation as Superradiant Transition (ST). 
We note that our $\kappa$ parameter bears some 
resemblance to the  Thouless parameter $g$, which in the case of opening 
at the edges of the system has been shown to provide relevant information on a
 metal-insulator transition \cite{Thouless}.
Here, the opening is obtained by coupling all sites to a common 
decay channel, and  {thus} $\kappa$ does not reveal the sensitivity to the
 boundaries of the system.

\begin{figure}[t!]
\centering
%\subfigure[]
%{\includegraphics[scale=0.15]{dickephasessuper}}
{\includegraphics[width=8cm,height=6.cm]{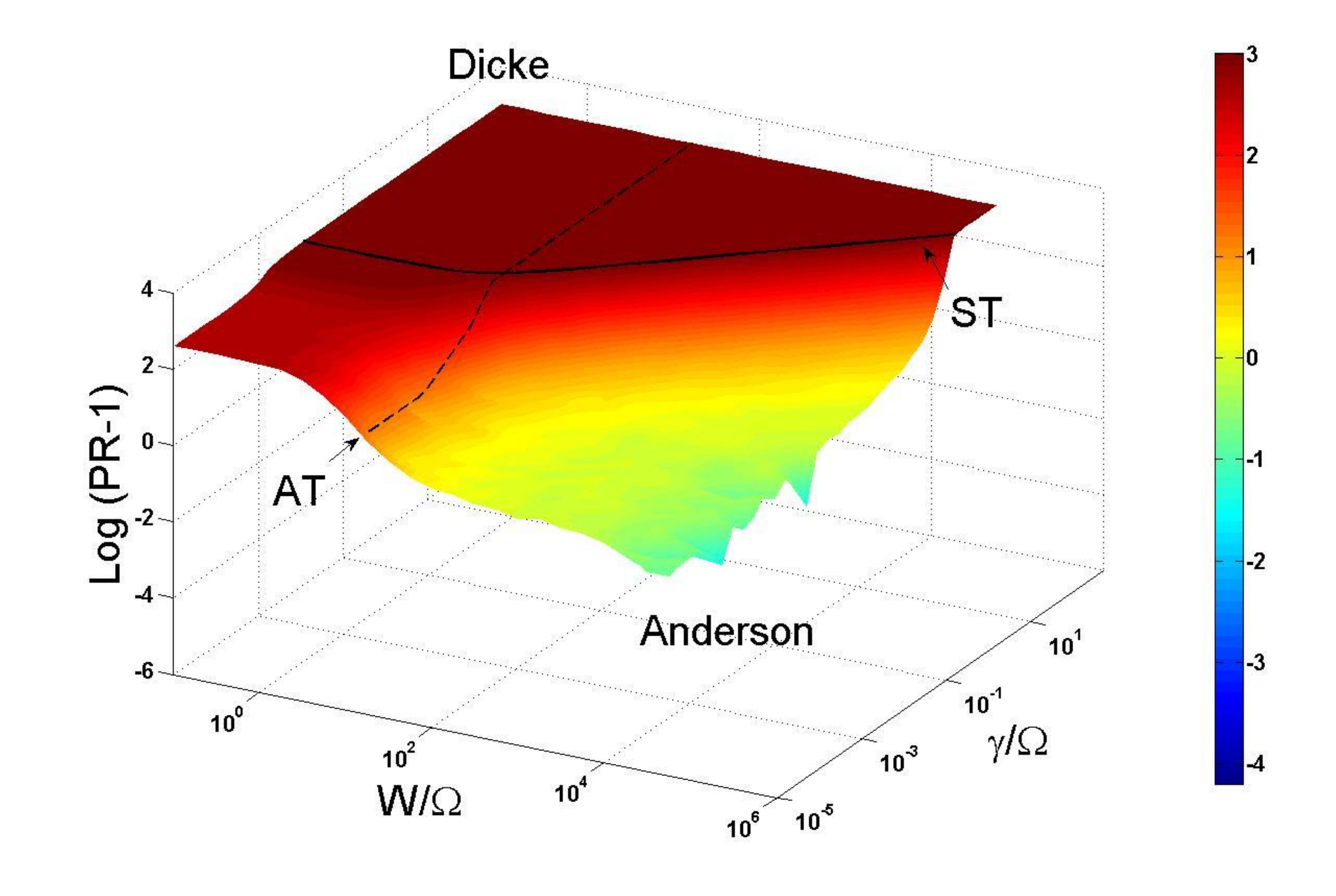}}

\hspace{1mm}
%\subfigure[]
%{\includegraphics[scale=0.15]{dickephases}}
{\includegraphics[width=8cm,height=6.cm]{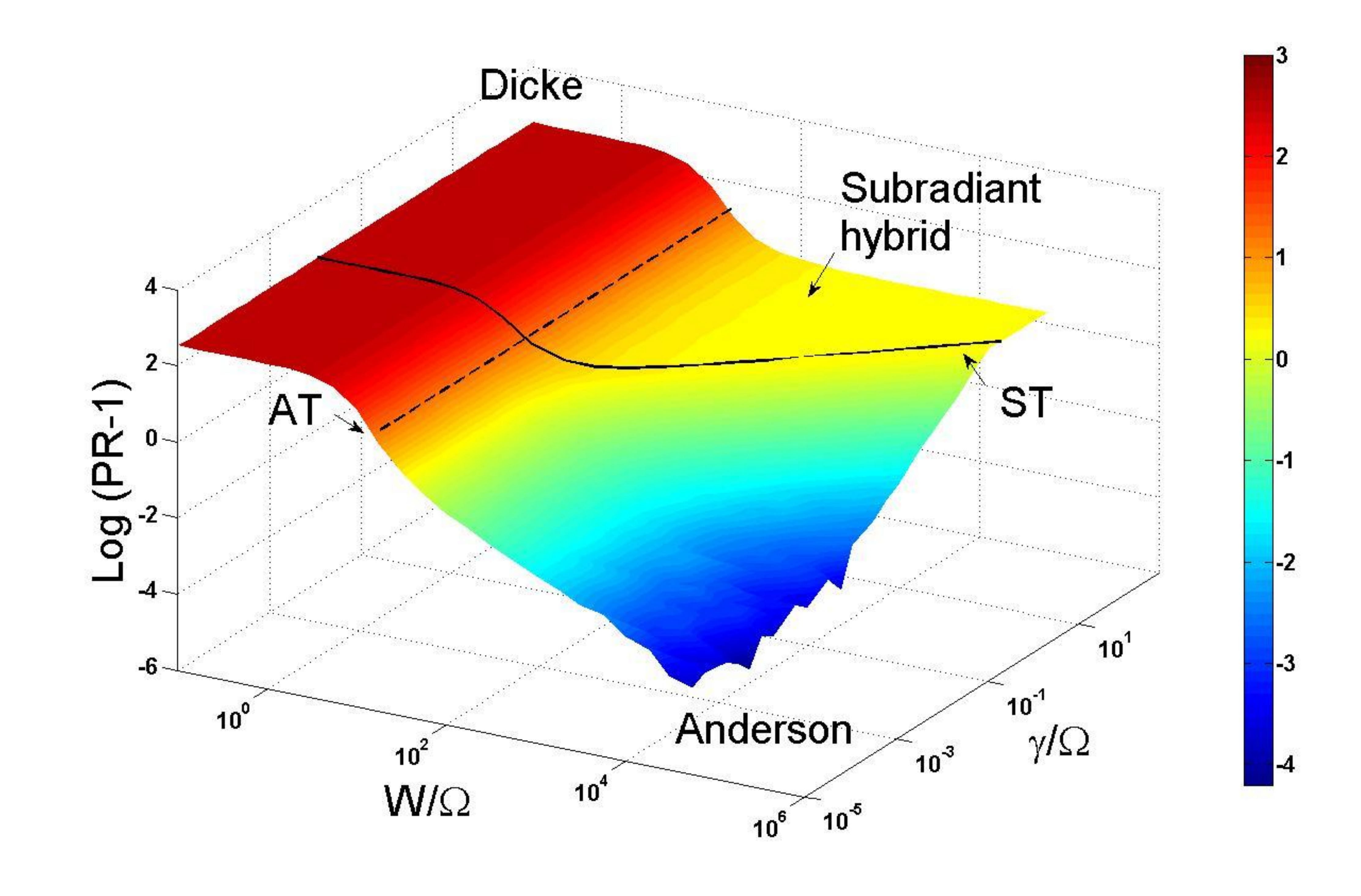}}
\caption{
Participation ratio as a function of both the 
opening ($\gamma/\Omega$) and disorder ($W/\Omega$).
Dashed line represents AT, Eq.(\ref{AT}), while
full line refers to the ST at $\kappa=1$, Eq.(\ref{3dkappa}).
Therefore
the region above ST, $\kappa > 1 $, is characterized by superradiant 
and subradiant states, differently from the region below ST, $\kappa < 1 $.
For an explanation of the different regimes see text.
Panel (a) refers to the state with the largest width, which 
become superradiant above the ST. Panel (b) refers 
to the other $N-1$ states which
become subradiant above the ST. 
In both panels the system is a cubic lattice with $N=10\times10\times10$ 
sites and $\Omega=1$. 
Each data is obtained averaging over $5$ realizations of disorder, for panel
(b) an additional average over all states has been done. }
\label{fig:phases}
\end{figure}

%\begin{figure}[t!]
%\includegraphics[scale=0.18]{dickephasessuper.eps}
%\includegraphics[scale=0.18]{dickephases.eps}
%\caption{Participation ratio as a function of both the 
%opening ($\gamma/\Omega$) and disorder ($W/\Omega$).
%Dashed line represents AT, Eq.(\ref{AT}), while
%full line refers to the ST at $\kappa=1$, Eq.(\ref{3dkappa}).
%Therefore
%the region above ST, $\kappa > 1 $, is characterized by superradiant 
%and subradiant states, differently from the region below ST, $\kappa < 1 $.
%For an explanation of the different regimes see text.
%Panel (a) refers to the state with the largest width, which 
%become superradiant above the ST. Panel (b) refers 
%to the other $N-1$ states which
%become subradiant above the ST. 
%In both panels the system is a cubic lattice with $N=10\times10\times10$ 
%%sites and $\Omega=1$. 
%Each data is obtained averaging over $5$ realizations of disorder, for panel
%(b) an additional average over all states has been done. }
%\label{fig:phases}
%\end{figure}

The presence of disorder affects ST since $D$ depends
on disorder. Indeed,
for small disorder, the width of the energy band is given by $\Omega$ and $D \simeq 12\Omega/N$~\cite{Anderson},
while for large disorder, $D \simeq W/N$, so that,
\
\begin{equation}
\begin{array}{cccc} 
ST &   {\rm at}  &  {\kappa = \frac{\gamma}{D}\simeq1}  &  { \rm with }
\begin{cases} 
\kappa \simeq \frac{\gamma N}{ 12 \Omega}  &  {\rm if }\,   \frac{W}{\Omega} \ll 12  \\ 
&\\
\kappa \simeq \frac{\gamma N}{W}   &  {\rm if }  \, \frac{W}{\Omega}  \gg 12  
\end{cases}
\end{array}
\label{3dkappa}
\end{equation}

So, above ST ($\kappa >1$), we have a superradiant regime, with superradiant and
subradiant states, while below ST ($\kappa <1$),
all states are affected by the opening in a similar way.  
From Eq.(\ref{3dkappa}) it is clear that we can control the 
effective degree of opening $\kappa$ for instance by varying the
strength of the disorder.

Since $Q$ is a full matrix 
the opening induces an effective long range hopping among the
sites, which is 
generally expected to destroy localization~\cite{levitov}. 
The long range hopping can be explained as an %second order 
effect of coupling via a common vacuum mode which results in 
the fact that the particle escaping from one site 
can be reabsorbed in  far  sites before leaving the system. 
This interaction mediated by the continuum, is at the origin of the ST.
Thus, disorder and opening have opposing effects:
while disorder tends to localize
the eigenfunctions, the opening tends to delocalize
them due to the induced  long range hopping.

\section{Results}

 {The numerical data presented in this letter were computed using the 
FORTRAN code (available at   http://www.netlib.org/eispack/cg.f),   which 
perform an exact diagonalization of $H_{\mathrm{eff}}$.
}
In order to analyze the interplay of disorder and opening we 
first study   the participation ratio, 
\beq
 PR= \langle [\sum_i |\langle i| \psi \rangle|^4 ]^{-1} \rangle, 
 \label{pr}
 \eeq
%$PR= \langle [\sum_i |\langle i| \psi \rangle|^4 ]^{-1} \rangle $
of the eigenstates $|\psi \rangle$  of $H_{\mathrm{eff}}$ 
in Eq.~(\ref{amef}),
where  $\langle \dots \rangle$  stands for the average over disorder {~\cite{average}}.
 {Since we want to study properties related to the probability densities, the eigenstates $|\psi \rangle$  of $H_{\mathrm{eff}}$ are normalized according
to $\sum_i |\langle i| \psi \rangle|^2$=1.
This  normalization differs from that needed 
to study the non-Hermitian time evolution, 
which should take into account the bi-orthogonality of the eigenstates. 
}
The $PR$ is widely used to characterize localization properties~\cite{pr}:
for extended states it increases proportionally to the system size,
while it is independent for localized states.
Note that in the following we will analyze $PR-1$.   

The interplay between disorder and opening
generates  {different regimes}, as shown in Fig. \ref{fig:phases},
where we plot $Log(PR-1)$ in the $\gamma/\Omega$--$W/\Omega$ plane
for a system with $N=10^3$ sites.
In the same figure  
 we also show  AT  (dashed curve) and 
ST ($\kappa=1$ full curve). 
In the upper panel of Fig. \ref{fig:phases} 
we analyze the state with the largest width which become superradiant
above ST,
while in the lower panel 
we consider the other $N-1$ states,  which become subradiant above ST.
 {As one can see
super- and subradiant states behave differently under the effect of disorder:
while the former
do not feel  AT and 
remain delocalized up to ST, the latter 
are sensitive to AT.  While this different
behavior can be described analytically using perturbation theory, see Eq.~(\ref{pertsubw}) below,
physically it can be explained by 
 the 
relative large distance of the SR state from the other states in the complex energy space, 
resulting in a weaker dependence on disorder~\cite{alberto}}.
%fact that the superradiant
%state is very far from the complex energies of the other states
%and thus it is less affected by the disorder~\cite{alberto}. 
Below the ST, all states
feel the disorder and the opening in a similar way.
From Fig. \ref{fig:phases} it is also clear that  
the parameters which determine
the nature of the eigenstates are  $W/\Omega$
and the effective coupling strength to the
continuum $\kappa$, Eq.(\ref{3dkappa}).
From Fig. \ref{fig:phases} we can distinguish   
three main regimes:
%\begin{itemize}
%\item 

\noindent
{\it Region I } (at the left of the AT line in  Fig. \ref{fig:phases}):
most of the states of the closed system are delocalized,
and the opening does not change their extended nature. 
For this reason we do not distinguish between
$\kappa<1$ and $\kappa>1$.
We  {avoid} the discussion about the effect of the 
opening on the mobility edges of the closed system,
since we are interested 
in
the global transition to localization  at the AT. 
Note that 
for large opening and small disorder we recover the usual Dicke regime. 

\noindent
%\item 
{\it Region II } (right of the AT line and below the ST line in  Fig. \ref{fig:phases} ): 
the states of the closed system
are localized and even if the effective opening is small
(we do not have superradiance), it
induces hybrid states (see discussion below).
It is important to note that this region disappears for large N values,
since, from Eq.~(\ref{3dkappa})
the critical ST line is, $\gamma/\Omega = 12/N$ for $ W/\Omega \ll 12 $ and
$\gamma/\Omega = W/(\Omega N)$ for $W/\Omega  \gg 12 $.
%\begin{equation}
%\begin{array}{cccc} 
%\gamma/\Omega =  &    &  &  
% \begin{cases}
%12/N &  {\rm for }\,   W/\Omega \ll 12  \\ 
%(W/(\Omega N)  &  {\rm for }  \, W/\Omega  \gg 12  
%\end{cases}
%\end{array}
%\label{3dkappag}
%\end{equation}
%and 
%Fig. \ref{fig:APD} (left panel). 
Note that in the limit of large disorder and small opening
we recover  the Anderson localized regime.

\begin{figure}[t!]
%\centering
%\includegraphics[scale=0.3]{APDW80}
\includegraphics[scale=0.4]{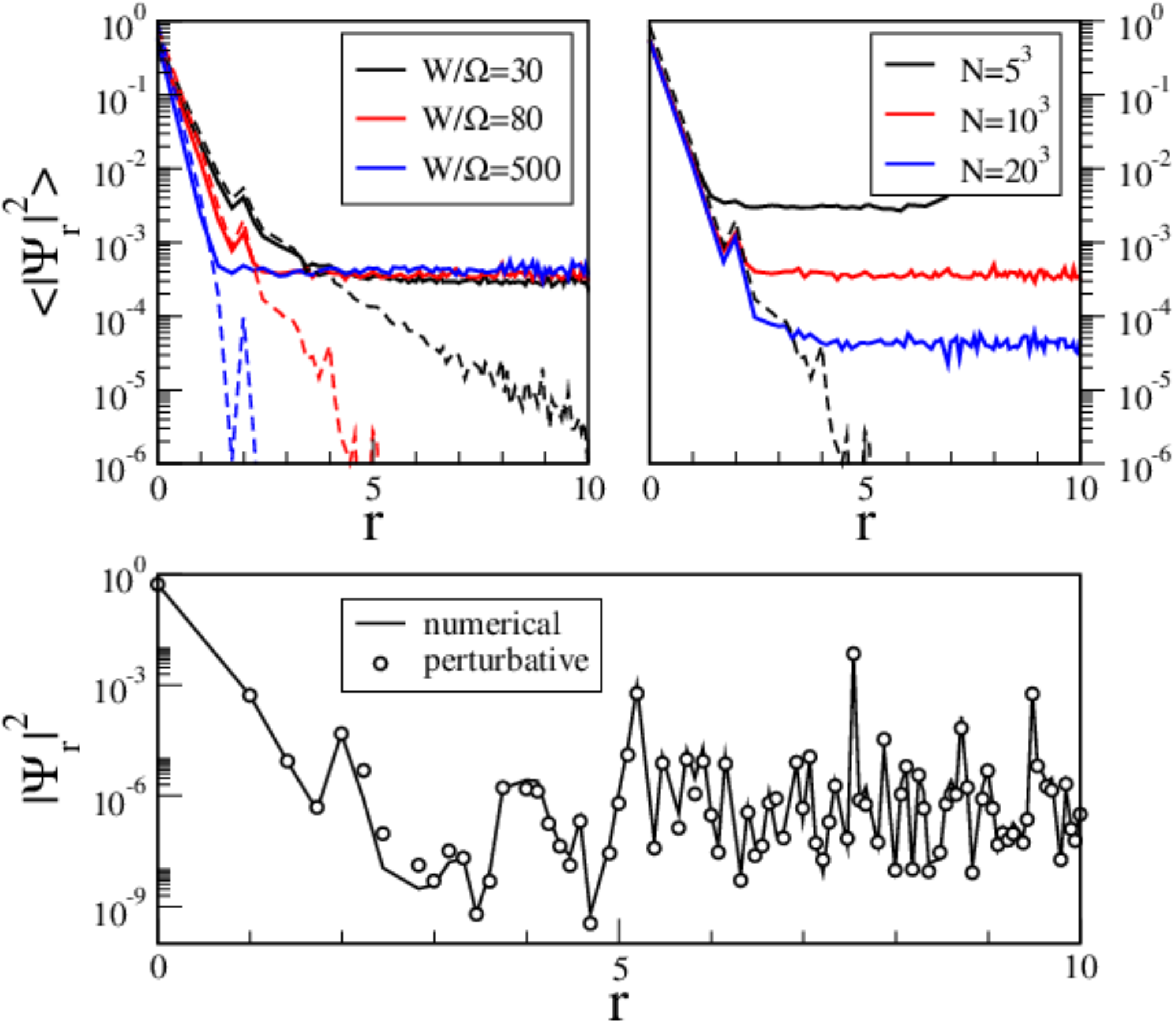}
\caption{
%\textcolor{red}{Average density profile of subradiant states
All data refer to the  {{\it subradiant hybrid regime}}.
%We chose different values of   $\gamma/\Omega$ and different $W/\Omega$, 
%so that we are in the {\it subradiant localized regime}. 
Full curves are for the open system, while the dashed curves 
are 
for the closed one.
%We chose strongly peaked states in the center of the cubic lattice.
In the two upper panels, the average density profile of subradiant states
as a function of the distance $r$ from the highest Anderson peak is shown.
Left upper panel: $N=10^3$, $\gamma/\Omega=100$ 
and different value of disorder as indicated in the legend.
%For the same disorder strength the corresponding density profile 
%of the states of the closed model are also shown (dashed line).  
Right upper panel:  $W/\Omega=80$, $\gamma/\Omega=1$  and  
different system sizes as indicated in the legend.
Lower panel: A single hybrid subradiant state is compared
with perturbation theory, Eq.(\ref{sub}).
 {Parameters are: $W/\Omega=80$, $\gamma/\Omega=1$ and $N=20^3$}. 
%In order to reduce the fluctuations we plotted 
%$\exp{(\langle |\psi|^2 \rangle)}$.  
%For each density profile an  average over $10$ disorder configurations
%was done.  
}
\label{fig:APD}
\end{figure}

\noindent
%\item 
{\it Region III or subradiant hybrid } (right of the AT line and above the ST line in  Fig. \ref{fig:phases} ): 
the states of the closed system
are localized, but the effective opening is large and
we have  superradiance. This regime is very interesting
since the superradiant state is fully delocalized 
(Fig. \ref{fig:phases}, upper panel), while
the subradiant states are hybrid 
 {with both localized and extended features}
%and localized 
(see discussion below and
Fig. \ref{fig:APD}).
 {Note that the nature of the hybrid states in this Region and in \mbox{\emph{Region II}} are different as explained below}.
The existence of this regime, which we 
call the  {{\it subradiant hybrid regime}}, 
is the main result of this work.  
%\end{itemize}

A full analysis of the three regions will be presented
in a future work. Here
we focus on the 
 {{\it subradiant hybrid regime}}, characterized by
both large opening and disorder, in which
one might expect no signature of localization.
%In the following we fix
%$\gamma/\Omega = 1 $ and vary  $W/\Omega$, so that
%the effective degree of opening $\kappa$, Eq.(\ref{3dkappa}) changes. 
%and the three different regimes defined above are crossed. 

\begin{figure}[t!]
\centering
\includegraphics[scale=0.45]{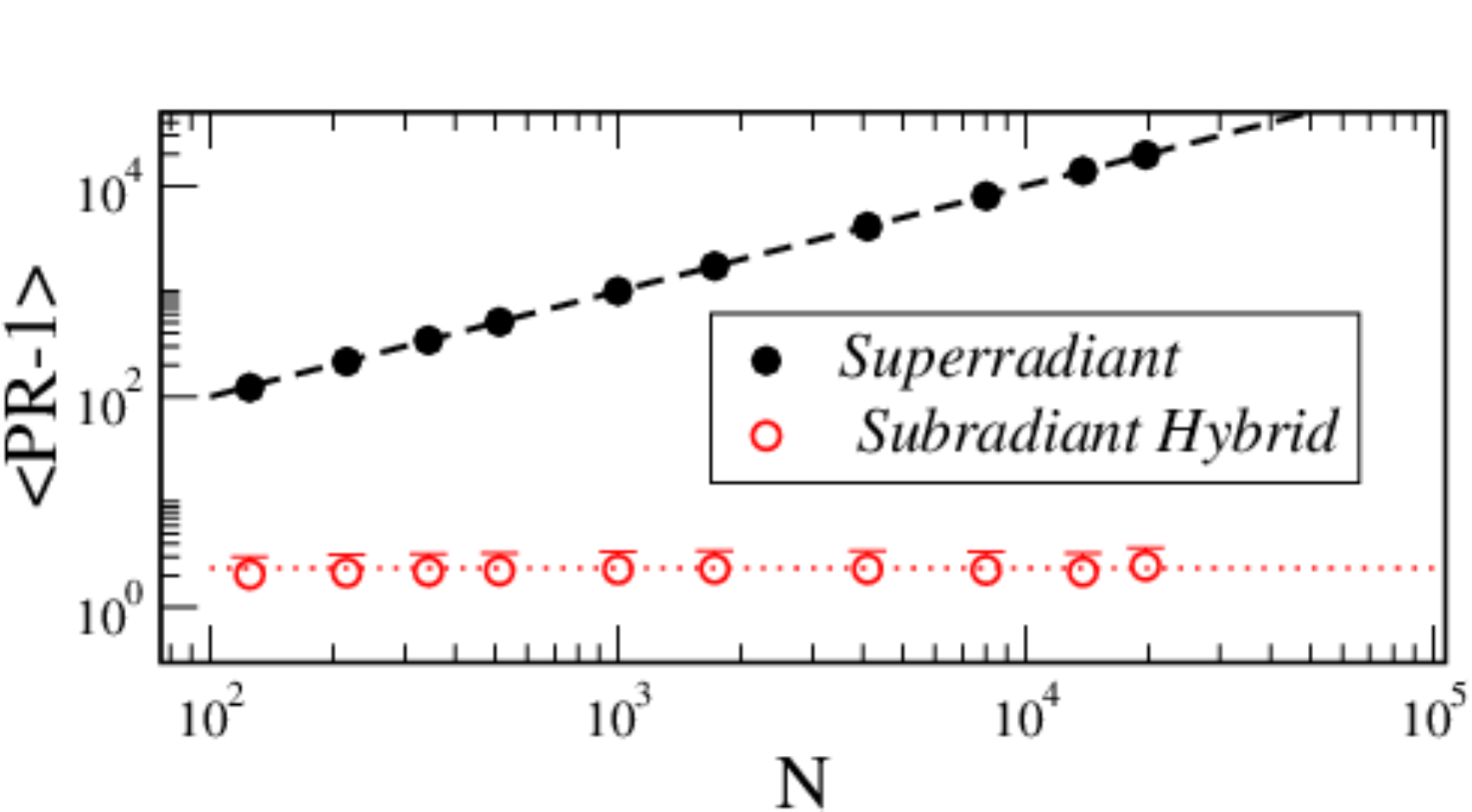}
\caption{$3D$ Anderson model. Average participation ratio $\langle PR -1 \rangle$   {\it vs}  $N$ (up to $N_{MAX}=27^3$).  Here  $\gamma/ \Omega= 1$ and
 $W/\Omega=80$, so that we are in the {\it Subradiant  {hybrid}  regime}.
Full circles stand for  the superradiant state, 
and  open red circles stand for  subradiant states.
Each point has been obtained by averaging both over the states and 
disorder.
Dashed lines are plotted to guide the eye. 
}
\label{fig:fscal}
\end{figure}

%Note that in Fig. \ref{fig:phases}, we fixed $N$, while
In order to establish the localized or extended nature
of the states, 
 {we start by analyzing} how the 
$PR$ scales with the system size. 
In  Fig. \ref{fig:fscal},
 $PR-1$ is shown {\it vs} $N$ 
for the subradiant and the superradiant states in 
the  {{\it subradiant hybrid regime}},
 at fixed strength of disorder.
As one can see the superradiant states (full circles)
are  extended, $PR \propto N$,
while, surprisingly, we find that the subradiant states behave
like localized states with a $PR$ independent of $N$
(open red circles).  
We have also checked the behavior of the $PR$ in other regions
(not shown):  Below AT  ({\it Region I}) we find that 
the states are extended ($PR \propto N$),
as in the closed model. 
In {\it Region II},  
we have found $PR \propto 1+ N$ which would indicated
extended states. 
 {
In reality, in the limit $N\to \infty$, 
the {\it Region II} disappears, since
$ W/\gamma \simeq N  $ and for large N
we cross the ST, entering the 
subradiant hybrid regime}.
We can conclude that this is a finite-size effect.
%despite the very small value of $PR$,  the states
%are extended ($PR \propto 1+ N$),
%differently from those of the closed system 
%which are localized since we are above AT.  
%Let us remark that this 
%region is a finite size region, since it disappears on increasing $N$. 
%Indeed at $N \simeq W/\gamma $ we cross the ST, entering the 
%{\it subradiant localized regime}.  

The behavior of the $PR$ is not enough to characterize the
nature of the eigenstates $\psi$, it is 
also important  to
analyze their structure,
 {for instance} by means of the average  density profile,
%measurable experimentally~\cite{aspect,bec} 
defined as $\langle |\psi|^2 \rangle$,
 { where the average $\langle \dots \rangle$ is taken
over different realizations of the disorder.}
In the  {{\it subradiant hybrid regime}}
the superradiant states are fully extended
and are well approximated by the following expression:
$$
|{\it Superradiant}\rangle \approx \frac{1}{N} \sum_i |i\rangle,
$$
where  $|i\rangle$ are the site states
of the Anderson model. 
The analysis of the subradiant states in the   {{\it subradiant hybrid regime}}
shows that
they are  hybrid~\cite{lagendijk}
with an exponentially localized
peak plus a constant plateau. 
The exponentially localized peak behaves like 
a localized state of the closed Anderson model
(compare respectively  solid  and dashed curves in Fig. \ref{fig:APD}).
On the other hand, the height of the constant plateau is independent of
both $W$ and $\gamma$, see  Fig. \ref{fig:APD} (left upper panel),
and it decreases with the system size as $1/N$, 
Fig. \ref{fig:APD} (right upper panel).
 {Thus the analysis of the structure of the subradiant states reveal their  hybrid
nature: even if their PR is size independent as for localized states, the
presence of a constant plateau 
make them also extended.
For instance,  due to the presence of the plateau, the transmission through the subradiant hybrid states will not decrease exponentially with the system size, as expected for
exponentially  localized states.
The study of the transmission through
hybrid state will be investigated in  a future work.}

For large disorder, $W \gg \Omega$ and above ST ({\it subradiant  {hybrid}  regime}), 
following Ref.~\cite{alberto, sokolov} 
it is possible to obtain an analytical expression for the  {hybrid} 
subradiant states:
\beq
|\mu\rangle =\frac{1}{\sqrt{C_\mu}} \sum_{j^0=1}^N \frac1{{\tilde{\epsilon}_\mu-E_{j^0}}} |j^0 \rangle\, \hspace {0.1cm} \mbox{with} \hspace {0.1cm} \sum_{j^0=1}^N\frac{1}{\tilde{\epsilon}_\mu-E_{j^0}}=0.
\label{sub}
\eeq
where $C_\mu$ is a normalization factor, 
$| j^0 \rangle$, $E_{j^0}$ are the eigenstates and eigenvalues of the closed Anderson model, Eq.(\ref{AM}),  and
$\tilde{\epsilon}_\mu$ are defined by the constraint given 
in the second equation of Eq.(\ref{sub}).
Note that  $\mu=1,..,N-1$ spans the subradiant subspace
and each $\tilde{\epsilon}_\mu$  lies between two neighbor
 levels $E_{j^0}$.
%\beq
%\sum_{j^0=1}^N\frac{1}{\tilde{\epsilon}_\mu-E_{j^0}}=0.
%\eeq
%where $\mu=1,..,N-1$ spans the subradiant subspace
%and each $\tilde{\epsilon}_\mu$  lies between two neighbor
% levels $E_{j^0}$.
Even if  Eq.(\ref{sub})
describes very well the  subradiant   {hybrid}  states, 
see  Fig. \ref{fig:APD} (lower panel),
it is not trivial to derive a closed expression for the density profile.
%Nevertheless we verified that 
%Eq.(\ref{sub}) is in very
%good agreement with the numerical results, see Fig. \ref{fig:APD} lower panel,
%where $|\psi|^2$ of a single wave function is compared
%with the perturbative results. 

%\begin{figure}[t!]
%\vspace{0.5cm}
%\centering
%\includegraphics[scale=0.4]{pertletter}
%\caption{The probability distribution of the $ln(PR)$ is plot in the 
%for $N=12^3$, $\gamma=\Omega=1$, $W/\Omega=80$, 
%for the subradiant states in the {\it subradiant localized regime}.
%Numerical results as a black line, while
%perturbative results are shown as a red line, see text. 
%}
%\label{fig:fpert}
%\end{figure} 

We have also computed the density profile of the
states for small opening and large disorder ({\it Region II}). 
When the  effective opening is small (below ST),
the problem can be treated by standard perturbation theory
and
similarly to what was found in~\cite{alberto},
we find that also in this case we have hybrid states,
with a  plateau height  
proportional to $(\gamma/W)^2$ but independent of $N$.
The different scaling of the plateau height  with
the system size, is at the origin of the different scaling of
the $PR$ of the open system between {\it Regions 
II} and the {\it subradiant  {hybrid}  regime}.
Indeed for very large
disorder, the hybrid states in both regions, are mainly 
localized on only one site with probability $a$ and 
on all the other $N-1$ sites with probability $b$, 
and we have $\sum_i|\psi_i|^4=a^2+b^2(N-1)$, 
with $a+b(N-1)=1$.
So that when $b \propto 1/N$ ({\it subradiant  {hybrid}  regime}), 
the PR %, Eq.(\ref{pr}),  
is independent
of $N$ in the large $N$ limit,
while, when $b$ is independent of  $N$
 ({\it Region II}) one has $PR \sim 1+2bN$ in the limit
 $bN \ll 1$, in agreement with the 
discussion given above.

\begin{figure}[t!]
\centering
\includegraphics[scale=0.38]{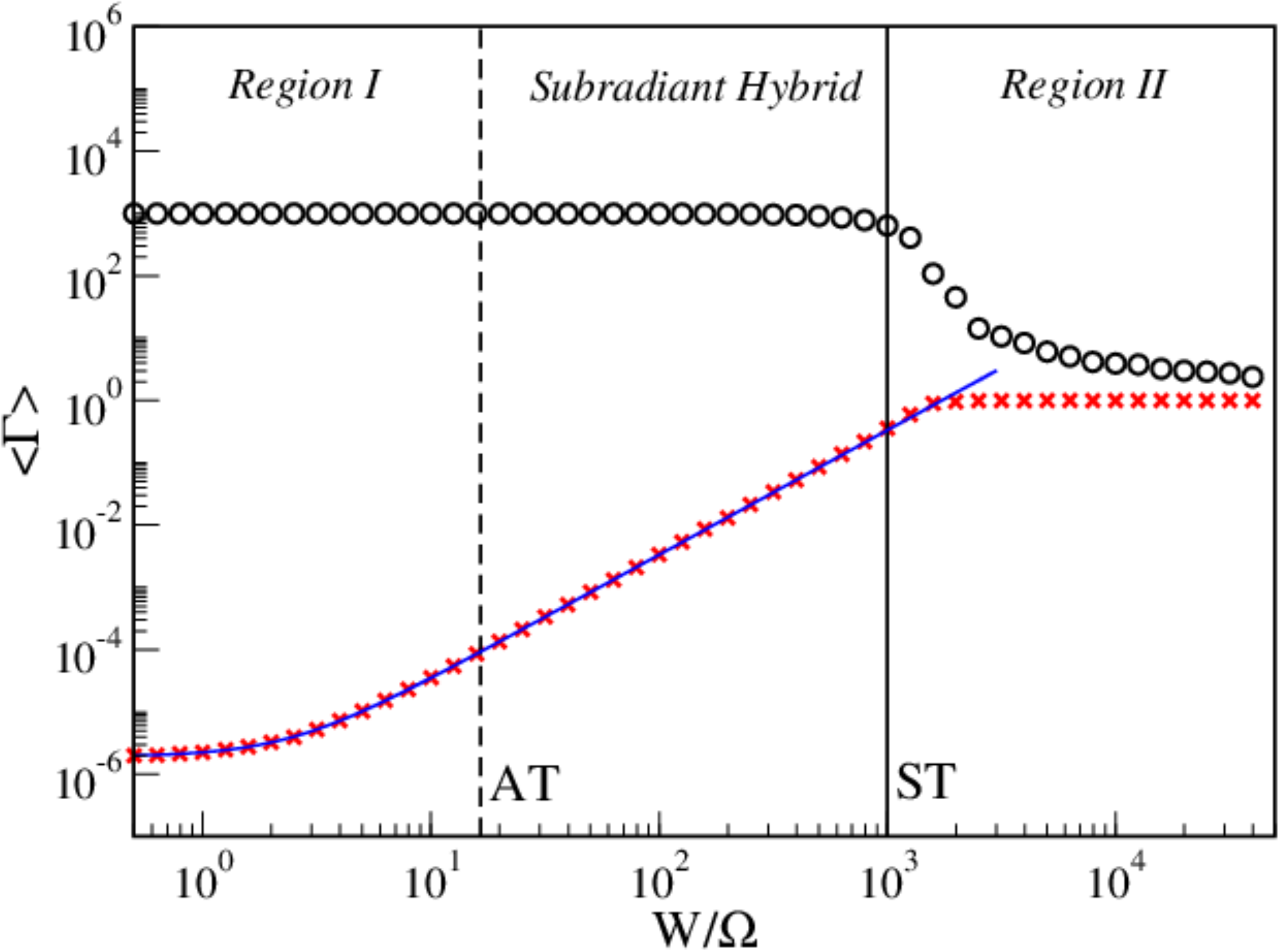}
\vspace{1.cm}
\caption{
Average width {\it vs}  $W/\Omega$ for $N=10^3$, and $\gamma = \Omega = 1$.
(Black) circles stand for the width of the superradiant state,
while (red) crosses stand for 
%the average widths of the 
%$N-1$ 
subradiant states. The vertical dashed line represents the 
AT, while the full vertical line
indicates the superradiant transition  (ST).
 {Each point has been obtained performing an arithmetic average both over the states and disorder}.
 {As a full curve we show the perturbative expression given in Eq.~\ref{pertsubw}.}
}
\label{fig:3dpaper1}
\end{figure}

A signature of the
AT can also be found  in the behavior of the
decay widths of the subradiant states.
In Fig.~\ref{fig:3dpaper1} we show 
the decay width of both the superradiant and
subradiant states as a function of the disorder.
While the former
is not affected by disorder up to the ST, the latter
feels the disorder above the AT
(note that, on increasing the disorder at fixed $\gamma$,  the effective degree of
opening is decreased and the three regions are crossed).
For very large disorder ({\it Region II}),  
all  widths approach the same value, $\gamma=1$,
which corresponds to the decay width of an isolated  site.
 {We also derived from perturbation theory in the small parameter $1/\kappa$ an analytical expression for the average width of the subradiant states 
\beq
\label{pertsubw}
\langle\Gamma\rangle(W)\simeq \langle\Gamma\rangle(W=0) + \frac{W^2}{3N^2\gamma},
\eeq
which is in good agreement with numerical results, see Fig.~\ref{fig:3dpaper1}}.

 \section{Conclusions }
We analyzed the coherent effects induced by the interplay of
coherent dissipation and disorder. 
For this purpose we considered 
a $3D$ open Anderson model in which a particle can
escape from any site to a common channel in the
continuum. This kind of opening induces a strong long range
hopping (all--to--all) between the sites of the Anderson model.
Contrary to expectations, we show that the opening
does not destroy all features of Anderson localization. 
Remarkably, 
for large effective opening, where we have superradiance,
we established the 
existence of a  {{\it subradiant hybrid regime}}
with extended superradiant states, and
 {hybrid} subradiant states 
(with a size independent participation ratio). 
We determined, both numerically and analytically, that
the subradiant states
have an hybrid nature,
with an exponentially
localized peak and a constant plateau.

This surprising  result
can be only explained by the
highly correlated nature of the long range hopping
present in this  system. 
These correlations induce a Superradiant Transition
where the coupling with the external world
is taken by the superradiant state, leaving the subradiant states
effectively decoupled, and thus able to preserve the
localized nature of the closed system.

Finally we note that the interplay between 
Superradiance and disorder has been also studied 
in a classical ensemble of interacting oscillators \cite{shah},
where the superradiant state has been shown to be resistant
to disorder in accordance with our findings.
Let us also observe that recently 
in  Ref.~\cite{ossipov},  {states with a size independent PR}, have been
found in a related model with  hermitian
long range hopping, thus supporting the generality
of our results.  
%our results
%agree with those found in a very recent paper by Ossipov,
%\cite{ossipov},  where
%a related model with  hermitian
%long range hopping was considered. The conclusions
%found in  Ref. \cite{ossipov} support the fact that  localization
%can exist also for hermitian long range hopping.

Our analysis is relevant
both from an applicative point of view, in the 
search of Anderson localization  in $3D$ dissipative systems 
(i.e. cold atoms), and from a
theoretical point of view (role of long range hopping
in the metal-insulator transition).

%Insert here the text.
%See fig.~\ref{fig.1}, table~\ref{tab.1} and eq.~(\ref{eq.1}).
%See also~\cite{b.a,b.b}.
%\begin{equation}
%\label{eq.1}
%0\neq1
%\end{equation}

%\begin{figure}
%\onefigure{epl-template.eps}
%\caption{Figure caption.}
%\label{fig.1}
%\end{figure}

%\begin{table}
%\caption{Table caption.}
%\label{tab.1}
%\begin{center}
%\begin{tabular}{lcr}
%first  & table & row\\
%second & table & row
%\end{tabular}
%\end{center}
%\end{table}

\acknowledgments
Support by Regione Lombardia and CILEA Consortium through a LISA Initiative
(grant 2011) and Universit\'a Cattolica (grant D.2.2 2011) is acknowledged.

\end{document}